\documentclass[%
twocolumn,
amsmath,
prstab,
]{revtex4-1}

\usepackage{graphicx}
\usepackage{hyperref}% add hypertext capabilities
\begin{document}

\title{Detection and clearing of trapped ions in the high current Cornell photoinjector}

\author{S. Full\thanks{sf345@cornell.edu}, A. Bartnik, I.V. Bazarov, J. Dobbins, B. Dunham, G.H. Hoffstaetter} 
\affiliation{CLASSE, Cornell University, Ithaca, New York 14853, USA}

\begin{abstract} 

We have recently performed experiments to test the effectiveness of three ion-clearing strategies in the Cornell high intensity photoinjector: DC clearing electrodes, bunch gaps, and beam shaking.  The photoinjector reaches a new regime of linac beam parameters where high CW beam currents lead to ion trapping.  Therefore ion mitigation strategies must be evaluated for this machine and other similar future high current linacs.  We have developed several techniques to directly measure the residual trapped ions.  Our two primary indicators of successful clearing are the amount of ion current removed by a DC clearing electrode, and the absence of bremsstrahlung radiation generated by beam-ion interactions.  Measurements were taken for an electron beam with an energy of 5 MeV and CW beam currents in the range of 1--20 mA.   Several theoretical models have been developed to explain our data.  Using them, we are able to estimate the clearing electrode voltage required for maximum ion clearing, the creation and clearing rates of the ions while employing bunch gaps, and the sinusoidal shaking frequency necessary for clearing via beam shaking.  In all cases, we achieve a maximum ion clearing of at least 70 percent or higher, and in some cases our data is consistent with full ion clearing. 

\end{abstract}

\maketitle

%%%%%%%%%%%%%%%%%%%%%%%%%%%%%%%%%%%%%%%%%%%%%%%%%
%
%  Introduction
%
%%%%%%%%%%%%%%%%%%%%%%%%%%%%%%%%%%%%%%%%%%%%%%%%%

\section{Introduction}

In an accelerator's vacuum chamber, any residual gas is rapidly ionized by collisions with the electron beam. At high beam currents, the resulting positive ions become trapped inside of the negatively charged beam and can cause a variety of effects including charge neutralization, coherent and incoherent tune shifts, optical errors, beam halo, beam losses, or even beam instabilities \cite{Georg05}.  Even with improvements in vacuum technology, ions can fully neutralize a beam within seconds for vacuum pressures as low as 1 nTorr.  Therefore one must directly remove the trapped ions to avoid or mitigate these potential effects.

Ion clearing methods are understood well enough to mitigate ion effects in most storage rings and synchrotrons \cite{poncet1, poncet2, poncet3}.  In low repetition rate linacs, ion trapping is typically not observed because the ions have time to drift out of the center of the beam pipe between bunches.  However, future linacs, such as the Cornell Energy Recovery Linac (ERL), enter a new CW high current regime where ion trapping is unavoidable  \cite{Georg05}.  Therefore it has become especially important to anticipate what clearing methods will be most effective in future high current linacs.

Up until this point, many simulations have been created to study ion trapping and clearing in ERLs \cite{Spethmann, Ati2}.  In general, these simulations are difficult to create and can have prohibitively long run times, particularly if one simulates space charge repulsion between a large number (upwards of millions) of ions.  Additionally, these simulations have yet to be verified by experiment, because data for this range of beam parameters is very scarce.  

The high current Cornell photoinjector is one of the few linacs in the world where different ion mitigation strategies can be experimentally tested for this parameter regime.  We have carried out a series of experiments to test the effectiveness of three different clearing methods: DC clearing electrodes, ion clearing bunch gaps, and resonant beam shaking.  

Clearing electrodes are essentially a parallel plate capacitor with an applied DC voltage in the range of 1 V to a few kV.  The electrodes are designed to overwhelm the electron beam's attractive potential and allow the ions to escape from the center of the beam pipe.  They are best employed in areas of high ion concentrations, which tend to be near beam size minima where ions accumulate due to longitudinal motion \cite{Spethmann}.  Although clearing electrodes can achieve significant ion clearing, often reducing the beam neutralization fraction to just a few percent \cite{poncet2, poncet3}, it is important to explore other clearing methods as well.  This is because the electrodes often benefit from the combined use of other techniques.  In addition, electrodes are often expensive to deploy around the recirculating path \cite{Georg05}.  This is especially true in large machines with relatively low beta functions (on the order of meters), as electrodes would need to be installed at most beam size minima.

In the second clearing method, short gaps between bunches are introduced to allow the ions to drift out of the center of the beam pipe while the beam is absent.  This technique has been employed in many electron storage rings with great success \cite{poncet2}.  In storage rings, this method is typically implemented by leaving a fraction of the ring empty at any given time.  In linacs and photoinjectors, this is done by turning the beam off for a given duration and at a certain frequency.  Unfortunately, several problems emerge when applying this method to ERLs and CW linacs.  The primary concern is that ERLs are particuarly susceptible to RF beam loading, one of the side effects of this technique \cite{Georg05}.  Secondary concerns involve beam instabilities such as the fast ion instability, which occur because ions still accumulate over the course of a single bunch train \cite{Raubenheimer1}.  One must choose the proper gap duration and frequency to achieve clearing while avoiding these other detrimental effects.

The third method, beam shaking, is a technique that was successfully employed in past accelerators such as the Anti-proton Accumulator Ring \cite{CernAA} and more recently in the Metrology Light Source (MLS) \cite{Feikes}.  This method involves applying a time varying voltage to a kicker or other electrode to transversely shake the beam and resonantly clear any trapped ions.  Even shaking amplitudes much smaller than the transverse beam size can result in significant reductions in trapped ion density.  Typically a single sinusoidal frequency is used, although broadband white noise has also shown to be effective \cite{Feikes}.  The necessary frequency is typically close to the ion oscillation frequency, and is usually determined experimentally by trial and error \cite{poncet2, Feikes}.

All three methods can result in significant clearing, and can be even more effective when deployed in tandem \cite{poncet1}.  In our experiments, we have examined each method independently to compare their effectiveness at clearing trapped ions, and have also developed several empirical models to explain and analyze our data.

%%%%%%%%%%%%%%%%%%%%%%%%%%%%%%%%%%%%%%%%%%%%%%%%%
%
%  Experiments
%
%%%%%%%%%%%%%%%%%%%%%%%%%%%%%%%%%%%%%%%%%%%%%%%%%

\section{Experiments}

Instead of measuring the effects of ions on the beam, we instead directly studied the trapped ions.  We chose to do this because the Cornell photoinjector is a relatively short accelerator, so any changes in beam dynamics due to ions may be difficult to observe directly.  Another contributing factor is that most traditional beam diagnostics are not viable in the photoinjector's parameter regime.  Due to the beam's high power at full current operation, any traditional interceptive beam diagnostics such as viewscreens, slits or wire scanners will quickly melt (with timescales typically on the order ms or lower). Additionally, because the photoinjector is a low energy linac, we are unable to use synchrotron or diffraction radiation to take measurements.  Our best option, a fast beam profile monitor recently developed at Cornell for use in high intensity accelerators \cite{wirescanner}, was unfortunately not available for use at the time of these experiments.

Larger machines, such as synchrotrons or storage rings, may observe the tune spectrum of a beam using beam position monitors (BPMs) connected to a spectrum analyzer \cite{Feikes, Wang}.  Ion-beam interactions lead to incoherent tune spreads and sidebands around the tune, so this is probably the easiest way to observe ion effects.  This measurement technique was attempted in the photoinjector, but no ion signatures were observed, even after leaking gas to increase the residual vacuum chamber pressure by a factor of 100.  This is likely because the small scale, non-circulating nature of the photoinjector means that beam-ion coupling must be visible on the spectrum analyzer after an interaction region of only about 6 m, which is simply too short.  

Instead we used two primary indicators of accumulated ions.  The first was a direct measurement of the trapped ion density using our clearing electrode.  By applying a DC voltage to the clearing electrode, the ions are drawn out of the center of the beam pipe, strike the clearing electrode and are measured by a picoammeter connected in series with the electrode.  The total ion current reaching the clearing electrode depends on the applied voltage, as will be shown below.  A sufficiently high voltage (in our case, only 28 V) will draw out all trapped ions in the vicinity of the clearing electrode.  

\begin{figure}[htb]
   \centering
        \includegraphics*[width = 85mm]{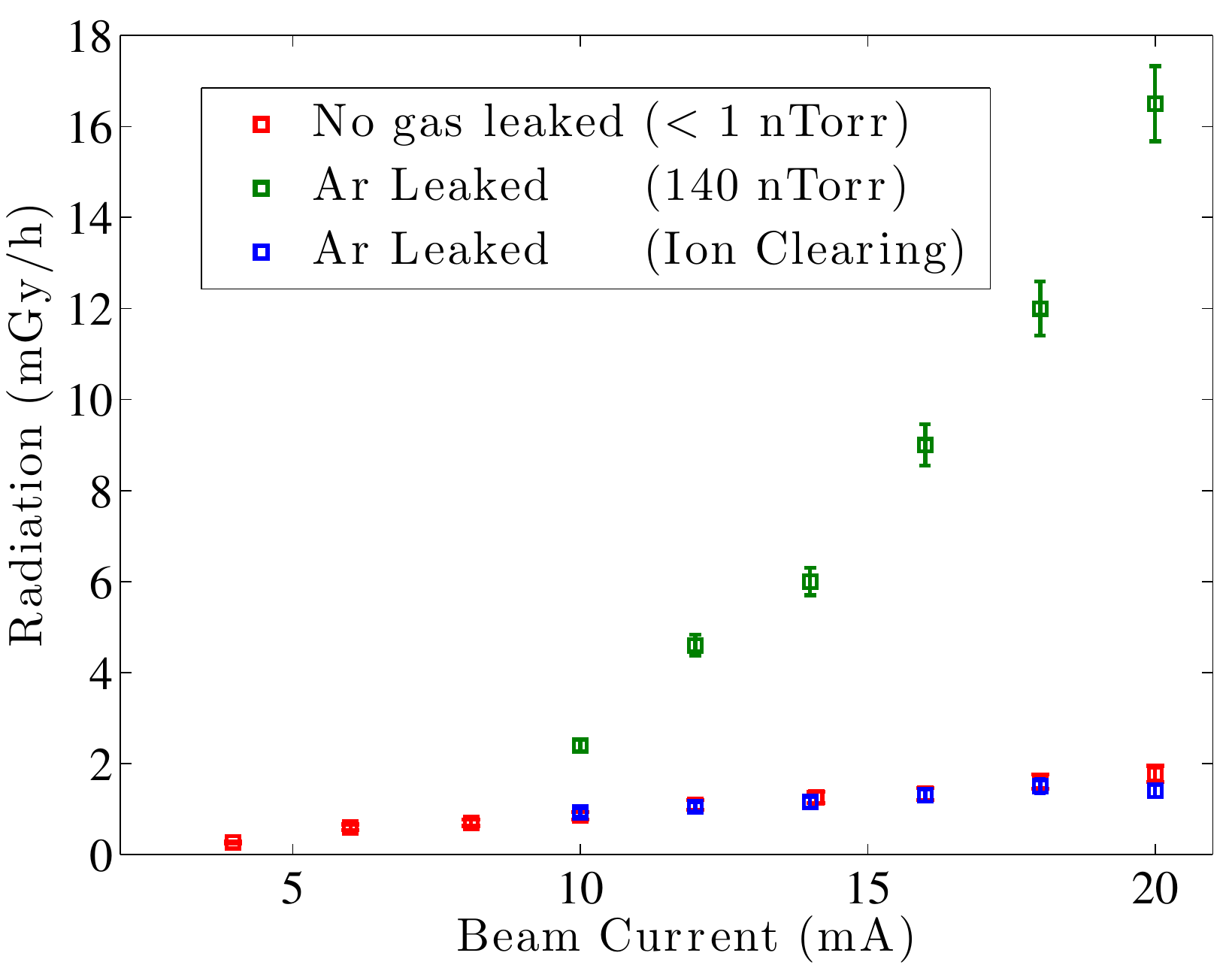}
   \caption{After leaking gas into the beam pipe, background radiation levels rose dramatically due to bremsstrahlung generated by beam-ion interactions.  Removing the trapped ions using clearing electrodes (shown above) or other clearing methods reduced this excess radiation to normal background levels.  }
   \label{measure_rad}
\end{figure}

We also used our radiation monitors as a secondary, indirect way of observing the trapped ion density.  The high power of the ERL photoinjector's beam generates large amounts of radiation, primarily created by beam losses and beam halo striking the beam pipe.  When the beam current was increased above 10 mA after gas injection, measured radiation levels rose sharply above normal background levels, as shown in Fig. \ref{measure_rad}.  Before leaking gas, no such excess radition was previously observed in the 10--20 mA range, indicating that this extra radiation (presumably bremsstrahlung) was caused entirely by beam-gas interactions.  All clearing methods significantly reduce this radiation, usually returning it to background levels.

\begin{figure*}[htb]
   \centering
   \includegraphics*[width = \textwidth]{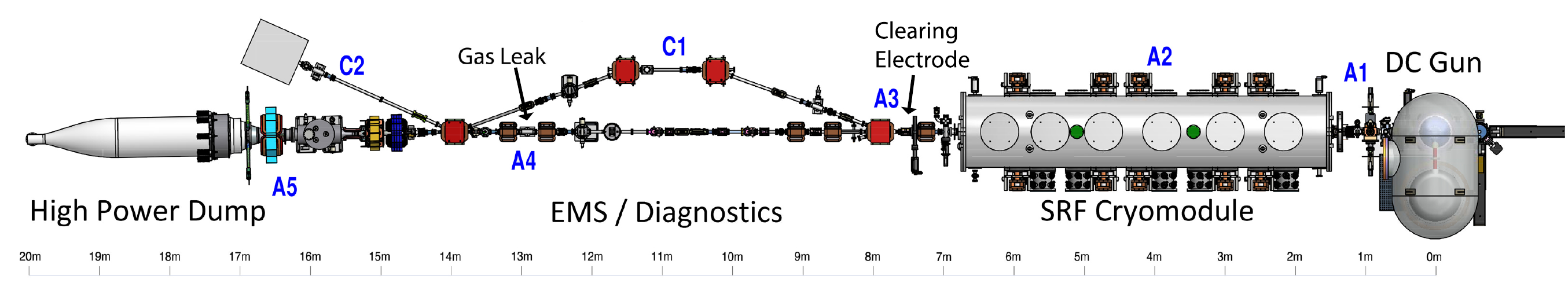}
   \caption{A schematic of the photoinjector.  The ion clearing electrode was installed just after the beam exists the SRF cavity at A3, and gas was leaked in at the end of A4.  Radiation measurements were taken at several locations between sections A3 and A4 (next to the beam pipe).}
   \label{fig_experimental_setup}
\end{figure*}

The experiments were performed in an approximately 8 m long straight section immediately after the beam exited the final accelerating cavity, as shown in Fig. \ref{fig_experimental_setup}.  Either $\text{N}_\text{2}$, Ar or Kr gas was leaked into the beam pipe so that the dominant ion species was known during the experiment.  The pressure in the beam pipe was increased to approximately 100--150 nTorr, as compared to typical values of 1--2 nTorr or less measured during normal operation.  

The photoinjector is designed to operate with a beam energy of 5--15 MeV and beam currents up to 100 mA, corresponding to a bunch charge of 77 pC at  a repetition rate of 1.3 GHz.  During these experiments we used a 5 MeV beam and varied the beam current from 1--20 mA by changing bunch charge.

%%%%%%%%%%%%%%%%%%%%%%%%%%%%%%%%%%%%%%%%%%%%%%%%%
%
%  Ion clearing electrodes
%
%%%%%%%%%%%%%%%%%%%%%%%%%%%%%%%%%%%%%%%%%%%%%%%%%

\subsection{Ion Clearing electrodes}

\begin{figure}[htb]
   \centering
   \includegraphics*[width = 85 mm]{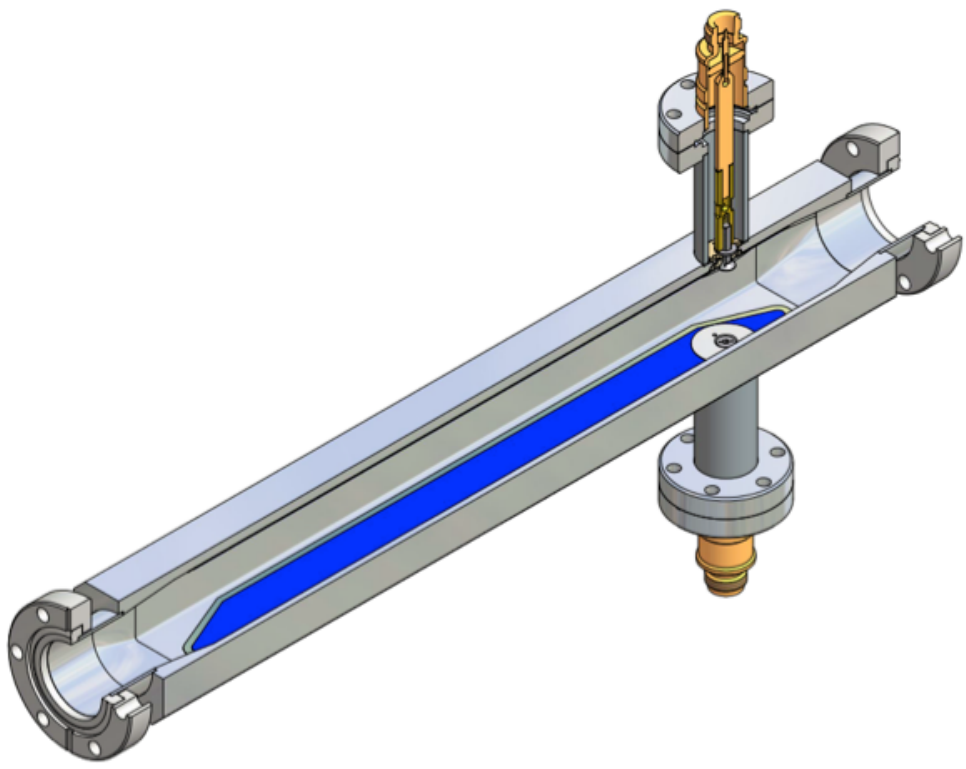}
   \caption{The ion clearing electrode used during the experiments.  The electrode coating is highlighted in blue, and is approximately 35 cm long and 3.5 cm wide.}
   \label{fig_clearing_electrode}
\end{figure}

Although it is possible to use button or stripline BPMs to clear ions, the photoinjector uses a specially created ion clearing electrode.  The device schematic is shown in Fig. \ref{fig_clearing_electrode} and its location in the beam line is shown in Fig. \ref{fig_experimental_setup}.  The electrode was oriented vertically during all experiments.  The clearing electrode surface is approximately 35 cm long and 3.5 cm wide, and it consists of two layers.  The bottom layer is a 0.30 mm thick alumina dielectric coating, and the top is a 0.20 mm thick tungsten electrode coating.  The top electrode was attached to a voltage supply, while the other was attached to ground.  The electrode's geometry is tapered to reduce wake fields, and it has been designed to allow for a maximum voltage of approximately 4 kV.  A picoammeter was attached in series with the voltage supply in order to measure the trapped ion current that was removed by the electrode.  

During this experiment we leaked $\text{N}_\text{2}$ gas into the beam vacuum chamber to raise the background pressure from a nominal value of less than 1 nTorr to 117 nTorr.  This ensured that we knew the dominant ion species present during the experiments.  After they are created via collision ionization, the ions drift longitudinally towards beam size minima.  This was taken into consideration when choosing beam optics for the experiment.

\begin{figure}[htb]
   \centering
   \includegraphics*[width=85mm]{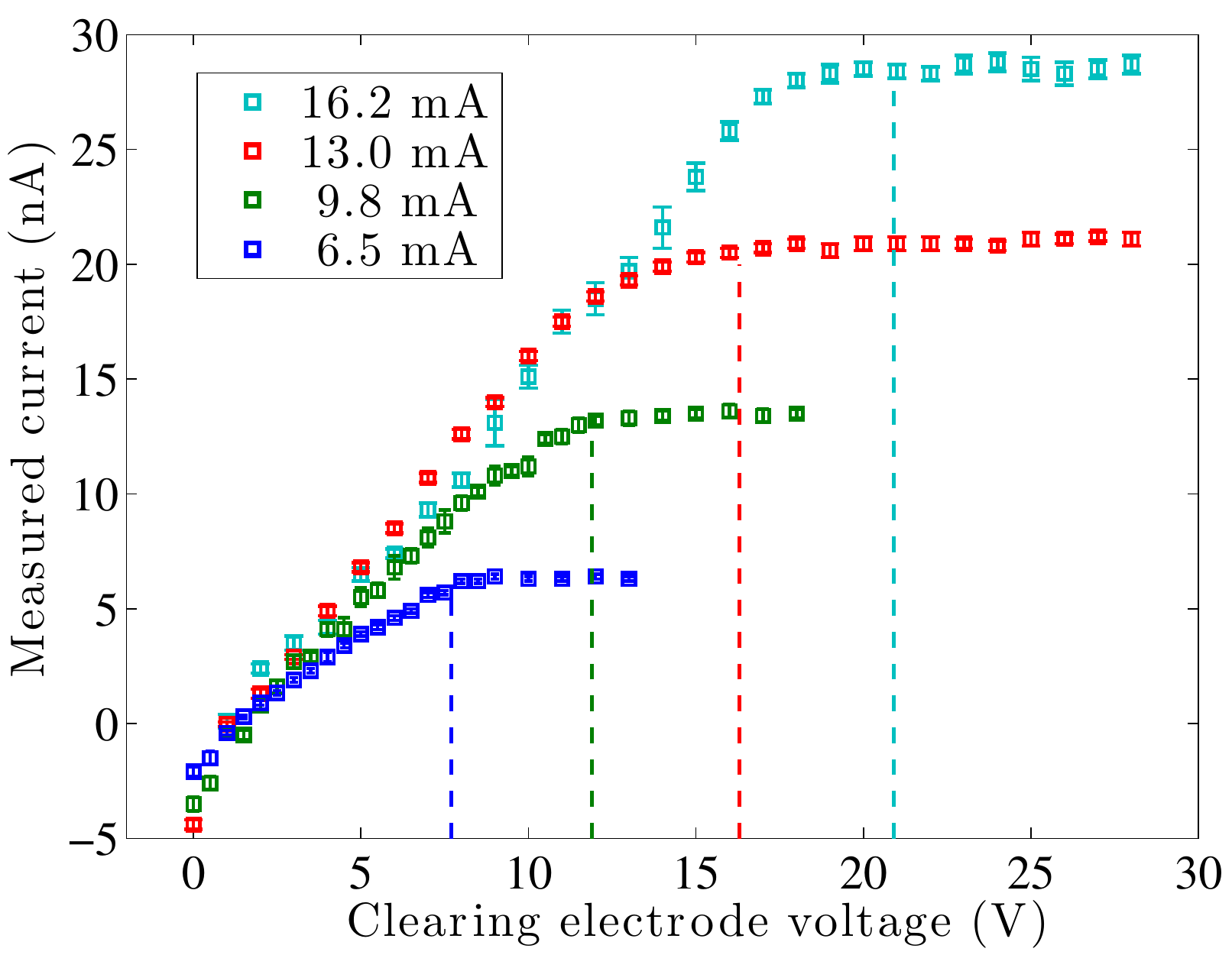}
   \caption{A picoammeter was used to measure the ion current striking the clearing electrode for different applied voltages.  The vertical dotted lines mark the minimum voltage required for full ion clearing, as predicted using equation (\ref{clearing_electrode_voltage}).}
   \label{figclearing_electrode_current}
\end{figure}

\begin{figure}[tb]
   \centering
   \includegraphics*[width=85mm]{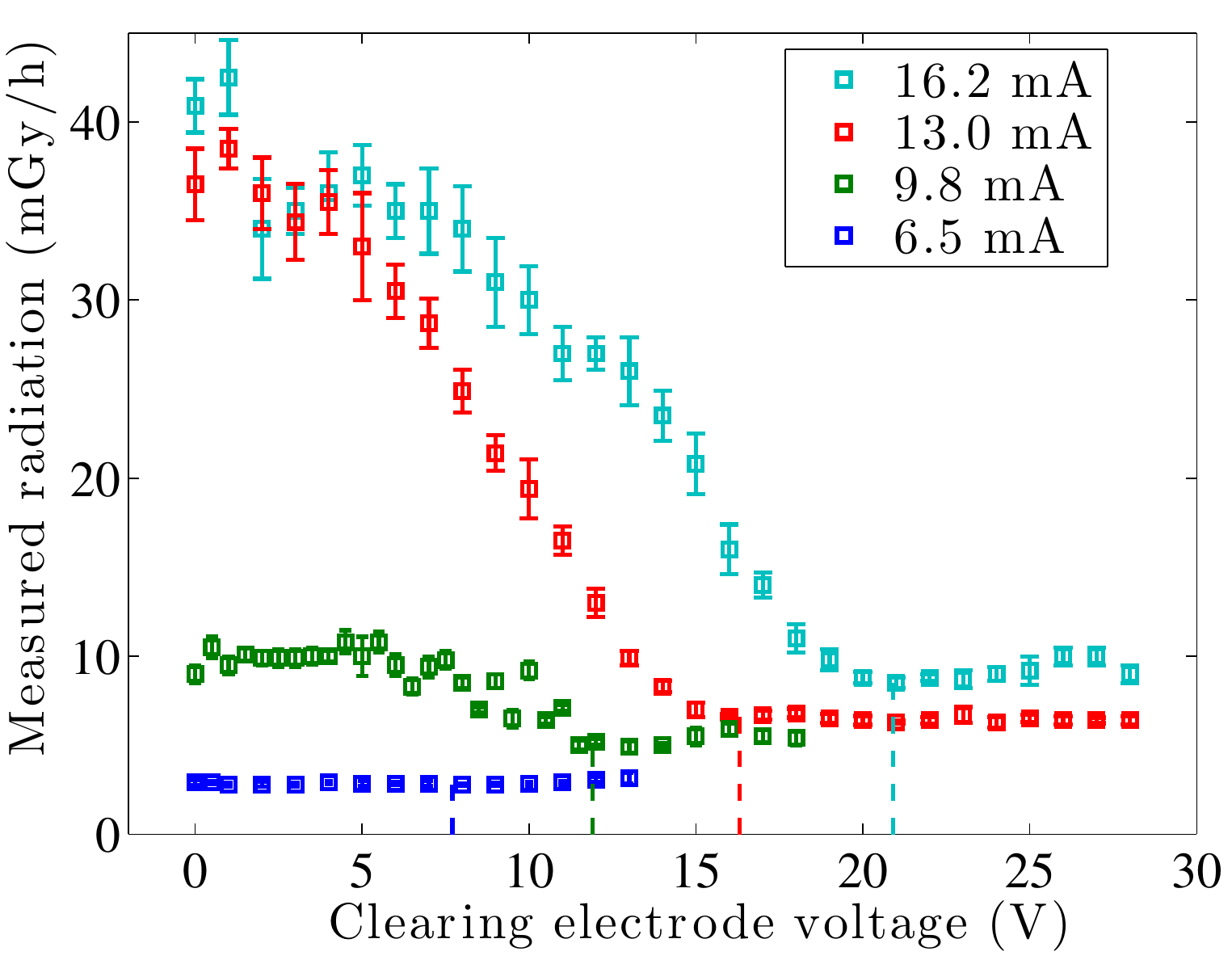}

   \caption{Background radiation levels also decreased while employing the clearing electrodes due to the absence of bremmstrahlung caused by beam-ion interactions.  The vertical lines again indicate the minimum voltage needed for complete ion clearing, as computed by equation (\ref{clearing_electrode_voltage}).}
   \label{figclearing_electrode_radiation}
\end{figure}

We varied the applied voltage on the clearing electrode from between 0 V and 28 V to test its effectiveness at clearing ions.  We looked at two signatures: the ion current striking the clearing electrode, and the background radiation observed by nearby radiation monitors.  Our data, taken for various beam currents between 5 mA and 20 mA, is shown in Fig. \ref{figclearing_electrode_current} and Fig. \ref{figclearing_electrode_radiation}.  The beam current was varied by changing bunch charge (from 5 pC to 12.5 pC) at a constant repetition rate of 1.3 GHz.  

Even when it was turned off, the clearing electrode measured a small background current.  The measured current was typically -4 nA (using the convention of a positive ion current).  This was true for both the clearing electrode and bunch gap experiments.  Also note that the measured radiation only exceeds background levels for beam currents greater than or equal to 10 mA, as can be seen in Fig. \ref{figclearing_electrode_radiation}.  This observation is typical throughout our experiments.  

The required voltage for maximum ion clearing can be predicted as follows:  Assuming a round beam of constant charge density, the beam's electric field is given by \cite{poncet3}

\begin{equation}\label{beam_electric_field}
	E_{beam}(r) = 
	\begin{cases}
		\frac{\lambda e}{2 \pi \epsilon_0} \frac{r}{\sigma_b^2}  , & \text{if } r \leq \sigma_b \\
    		\frac{\lambda e}{2 \pi \epsilon_0} \frac{1}{r} ,              & \text{if } r \geq \sigma_b
	\end{cases}
\end{equation} 
\noindent

\noindent
where $\lambda$ is the number of electrons per unit length, $e$ is the elementary charge, and $\sigma_b$ is the rms transverse beam size.  Full clearing occurs when the clearing electrode's field exceeds the beam's peak electric field (at $r = \sigma_b$).  For electrodes separated by a distance $d$, the field of the electrode is given by $E = \frac{V}{d}$, as the clearing electrode is very nearly a parallel plate capacitor.  Therefore we examine the case where $E_{electrode} = V_{electrode} / d \geq E_{beam}$.  This yields the minimum voltage required

\begin{equation}\label{clearing_electrode_voltage}
	V_{electrode} \geq \frac{\lambda e}{2 \pi \epsilon_0} \frac{d}{\sigma_b}
\end{equation} 

\noindent
Although we cannot take direct beam profile measurements in order to determine $\sigma_b$, we can obtain estimates using General Particle Tracer (GPT) \cite{Colwyn}, a 3D space charge simulation code that models the photoinjector.  This simulation is found to be in good agreement with results at low beam currents \cite{Colwyn}, although it has yet to be experimentally verified for beam currents above 1 mA.  For now we assume that the ions or other high current effects do not change the beam size significantly.  Using the vertical beam size from GPT and a clearing electrode separation of $d = 40.6$ mm, equation (\ref{clearing_electrode_voltage}) predicts clearing voltages that agree with measurements to within a few percent, as shown in Fig. \ref{figclearing_electrode_current} and Fig. \ref{figclearing_electrode_radiation}.  Our calculated values are shown in Table 1.  Vertical dotted lines were drawn in Fig. \ref{figclearing_electrode_current} and Fig. \ref{figclearing_electrode_radiation} to guide the eye and make it easier to compare these calculated values with our data.

The maximum measured ion current can be used to obtain an estimate of the longitudinal range of the clearing electrode.  In the absence of clearing, ions will accumulate (via collision ionization) until the total number of ions per unit length equals the total number of beam electrons per unit length.   The time it takes to accumulate as many ions as electrons per unit length is given by \cite{Georg05}

\begin{equation}\label{neutralization_fraction1}
	\tau_{create} =  \frac{1}{\sigma_{col} \rho_{gas} c}
\end{equation} 

\noindent
where $\sigma_{col}$ is the collision ionization cross section of the gas species (these values are readily available \cite{ionization_cross_sections}), $\rho_{gas}$ is the residual gas pressure, and $c$ is the speed of light.  

\begin{figure}[htb floatfix]
   \centering
          \includegraphics*[width=85mm]{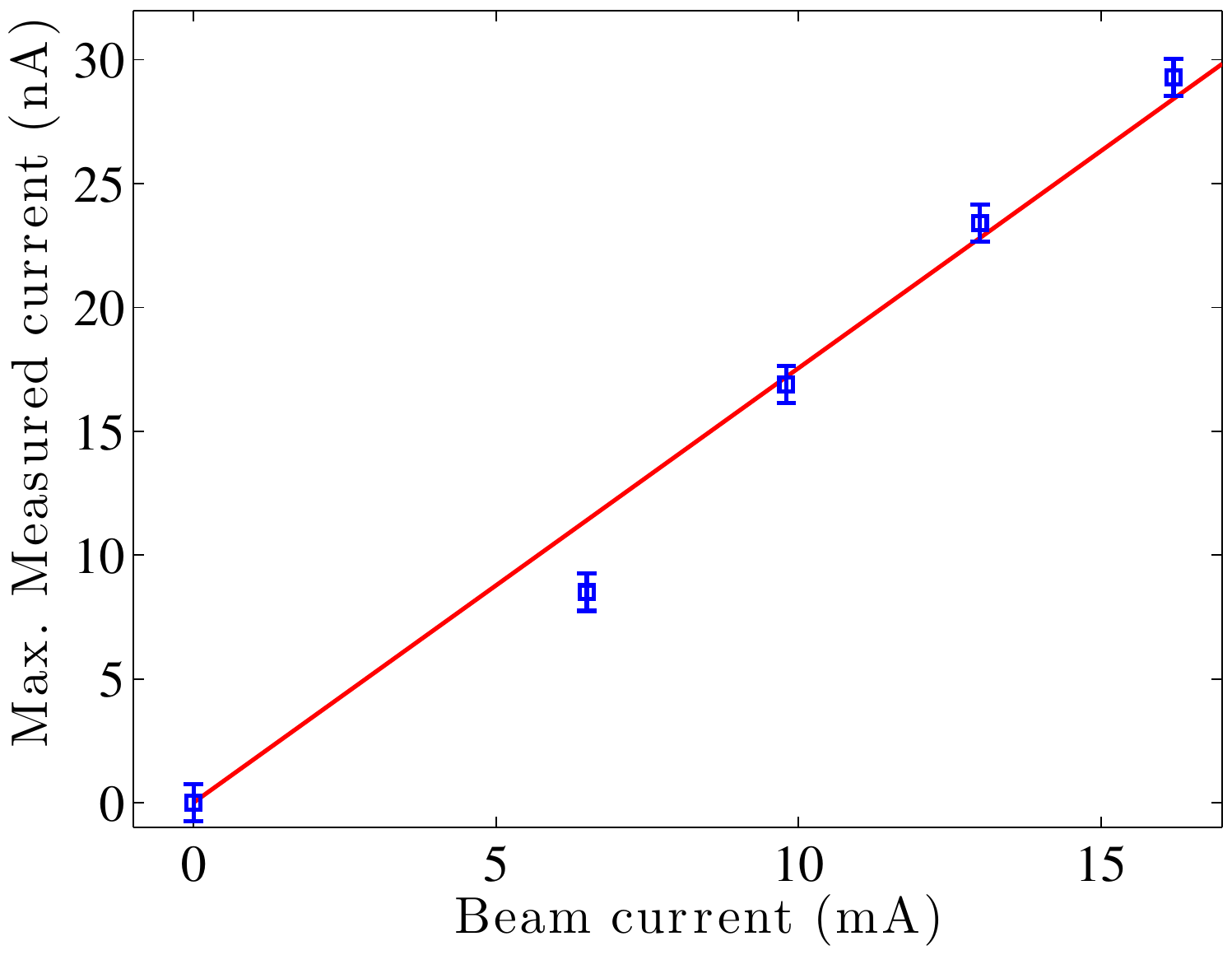}
   \caption{The longitudinal range of the clearing electrode can be estimated by using the maximum ion current measured by the clearing electrode.  The range can be extracted from a best fit line (in red) to our data.  }
   \label{fig_long_range_estimate}
\end{figure}

\begin{table}[b floatfix]
\label{table_voltage_calculation}
\caption{The minimum clearing electrode voltage necessary for full ion clearing, calculated using equation (\ref{clearing_electrode_voltage}).  The rms transverse beam sizes $\sigma_x$ and $\sigma_y$ were obtained using GPT simulations \cite{Colwyn}.}
\begin{tabular}{cccc}
	\toprule
	\textbf{Current (mA)             } & \textbf{     \boldmath$\sigma_x$(mm)              } & \textbf{     \boldmath$\sigma_y$(mm)              } & \textbf{     Voltage (V)} \\
    	\hline
	6.5   &  2.11 & 2.05 &  7.7 \\
	9.8   &  2.07 & 2.00 & 11.9 \\
	13.0 &  2.03 & 1.94 & 16.3 \\
	16.2 &  1.98 & 1.89 & 20.9 \\
    	\botrule
\end{tabular}
\end{table}

Over a longitudinal region $L$, there are $\lambda L$ beam electrons.  We define a region $L_{create}$ such that all of the ions over this region are removed by the clearing electrode.  On average, an electron needs the time $\tau_{create}$ to produce one ion. In the length L, each electron produces $L/c\tau_{create}$ ions. The beam current therefore produces $ L/c\tau_{create}\cdot I/e $ ions per second in this section. Thus the maximum ion current is 

%Assuming that the beam is fully neutralized, the number of removed ions should equal the number of electrons $\lambda L_{create}$.  At equilibrium, the number of ions created per second ($\lambda L_{create} / \tau_{create})$ will equal the number of ions removed per second.  Therefore the maximum removed ion current is given by 

\begin{equation}
\label{clearing_electrode_current}
%	I_{max} =  \lambda e \bigg(\frac{ L_{create} }{\tau_{create}}\bigg)
	I_{max} =  \frac{I_{beam}}{c} \bigg(\frac{ L_{create} }{\tau_{create}}\bigg)
\end{equation} 

\noindent
The ratio $L_{create}/\tau_{create}$ can be found by treating it as a fit parameter to our data, shown in Fig. \ref{fig_long_range_estimate}.  Note that in Fig. \ref{fig_long_range_estimate} we have adjusted our data to take into account background levels.

If the value for $\tau_{create}$ is known, then we can obtain an estimate for the longitudinal range $L_{create}$.  For a $\text{N}_\text{2}$ gas pressure of 117 nTorr and an assumed temperature of 300K, this corresponds to a beam neutralization time $\tau_{create}$ of roughly 5.2 ms.  Therefore, using the fit parameter found in Fig. \ref{fig_long_range_estimate}, the creation length $L_{create}$ should be roughly 2.7 m.  This estimate seems reasonable, considering that the distance between the gas leak and the clearing electrode is roughly 5 m.  In reality, it is possible that the longitudinal range will increase with the applied clearing voltage.  In the future, taking data points for more beam currents should allow us to find a more accurate estimate.  

%%%%%%%%%%%%%%%%%%%%%%%%%%%%%%%%%%%%%%%%%%%%%%%%%
%
%  Bunch Gaps
%
%%%%%%%%%%%%%%%%%%%%%%%%%%%%%%%%%%%%%%%%%%%%%%%%%

\subsection{Bunch Gaps}

While storage rings can create gaps simply by leaving a fraction of the ring empty at any given time, CW linacs require the introduction of a short bunch gap every few milliseconds.  In our experiments this was achieved by using a Pockels cell (normally used to select pulses for our low repetition rate emittance measurements \cite{Colwyn}) to reject laser pulses with a given duration and at a certain frequency.  Due to the hardware limitations of our Pockels cell, we were unable to create a bunch gap larger than 10 $\mu s$.  This experiment was performed immediately following the ion clearing electrode experiments, so the gas pressure remained at 117 nTorr for $\text{N}_\text{2}$.   

During these gaps, the regulation of the fields in the SRF and buncher cavities struggled to handle the sudden change in beam loading. As the beam current was increased, the error in the field amplitude and phase increased until they reached their pre-defined limits and tripped off the machine at around 8 mA.  We have a pre-existing feedforward system \cite{Florian}, originally designed to handle the analogous situation when there are short bunch trains for emittance measurements. This system was able to completely remove the amplitude and phase errors in the SRF cavities without any modifications.  However, for the buncher, the feedforward became unstable above a certain amount of gain.  At the time of the experiment, we decided to just limit the gain rather than investigate the cause of the instability. As a result, we were limited in beam current to around 20 mA by the remaining phase error in the buncher.  Ultimately, we believe that this is not a fundamental limitation to this bunch gap clearing method.  With more work, we believe we could solve this problem in the future.  

\begin{figure}[t floatfix]
   \centering
      \includegraphics*[width=85mm]{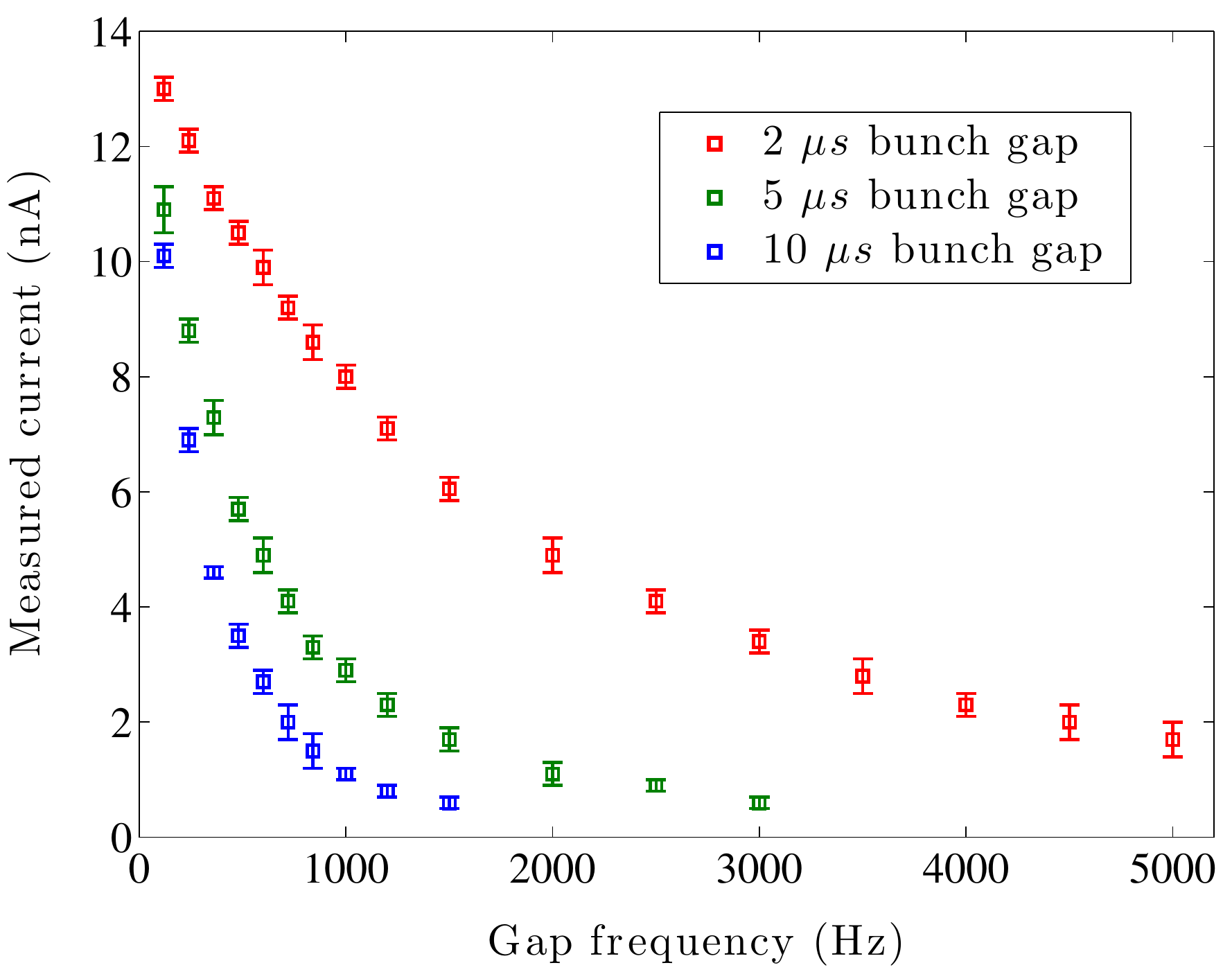}
   \caption{Increasing the frequency and duration of bunch gaps reduces the number of trapped ions that reach the clearing electrode.  For each data point, the beam current was held fixed at 10 mA.}
   \label{figbunch_gaps1}
\end{figure}

\begin{figure}[t floatfix]
   \centering
      \includegraphics*[width=85mm]{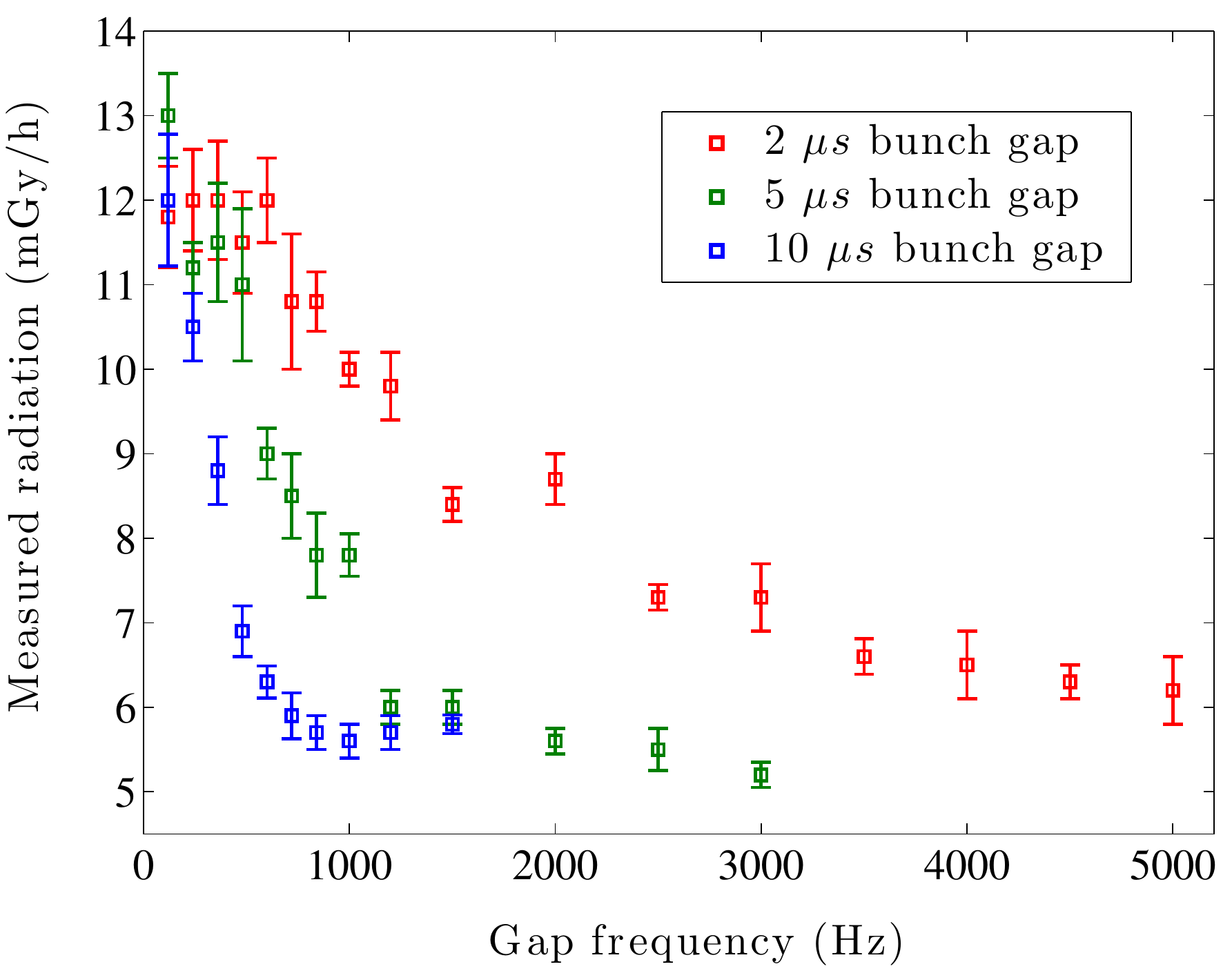}
   \caption{The radiation caused by beam-ion interactions is reduced by increasing the frequency and duration of bunch gaps.  This data was obtained while the clearing electrode was turned off.}
   \label{figbunch_gaps3}
\end{figure}

When employing bunch gaps, a fraction of the trapped ions drift transversely out of the beam during the gaps and into the vacuum chamber walls.  The remaining trapped ions travel longitudinally down the beam pipe towards our clearing electrode and are measured by the picoammeter.  We applied a large enough voltage (28 V) to the clearing electrode to ensure maximum ion clearing.  Thus we are measuring the amount of ions that remain trapped in the beam after clearing via bunch gaps.  Data for an average beam current of 10 mA was taken for various bunch gap lengths and frequencies, and is shown in Fig. \ref{figbunch_gaps1} -- \ref{figbunch_gaps3}.  The radiation data in Fig. \ref{figbunch_gaps3} shows that the trapped ions are removed even without the clearing electrode turned on, confirming that the bunch gaps are the dominant clearing mechanism.

\begin{figure}[htb]
   \centering
   \includegraphics*[width=85mm]{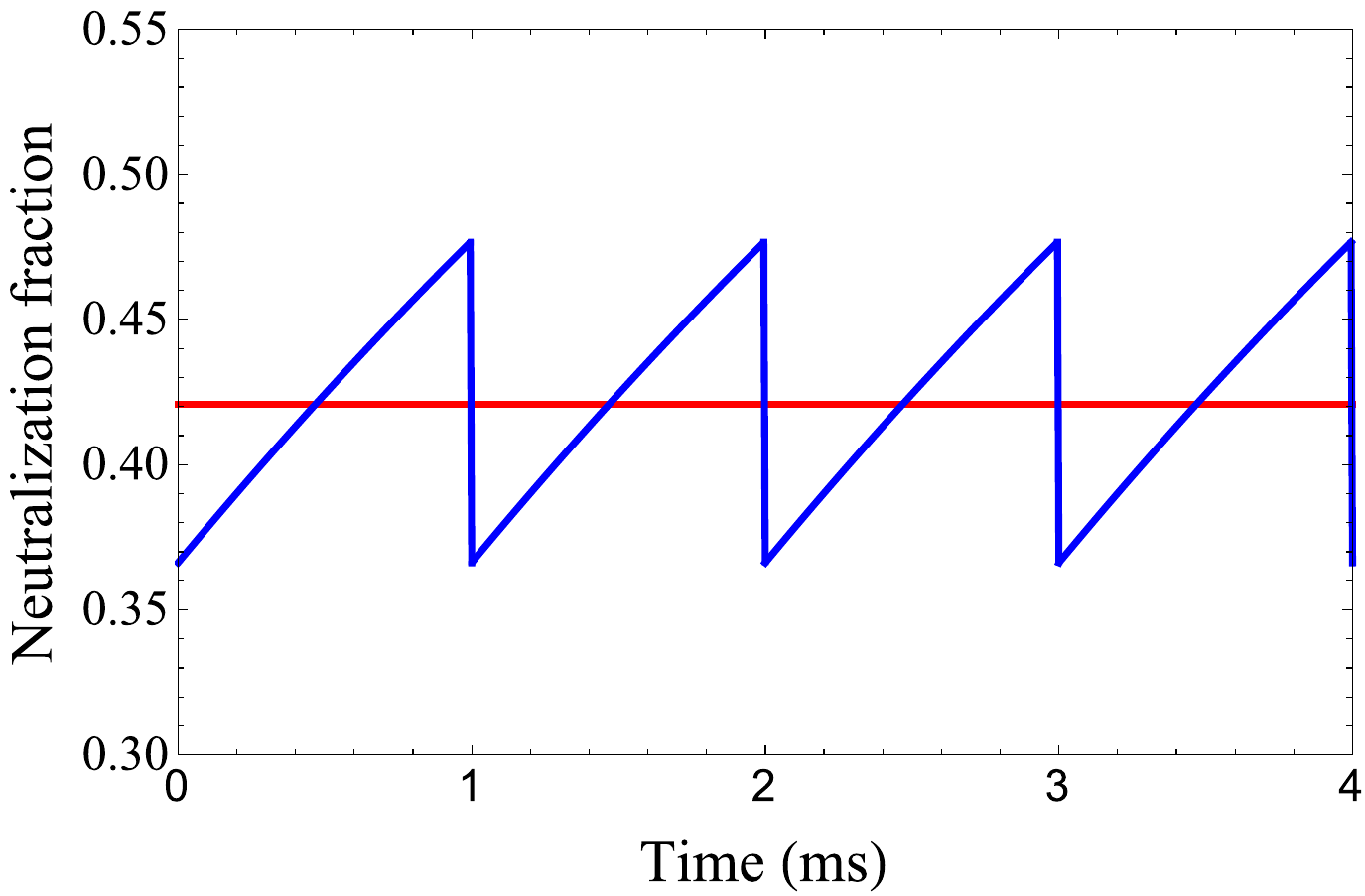}
   \caption{Ions are created via collision ionization while the beam is on and decay exponentially during the bunch gaps.  The equilibrium neutralization fraction, indicated by the red line, was found using equation (\ref{ridiculously_simplified_fraction}) for a bunch gap duration of 5 $\mu s$, a gap frequency of 1 kHz, and the ion creation to clearing ratio given in Table \ref{table_ratio}.   }
   \label{figsawtooth}
\end{figure}

For now we have devised an empirical model to explain our data.  In our simple model we assume that while the beam is on, ions are created via collision ionization.  The neutralization fraction should never exceed 1, because when the beam is fully neutralized, its potential well will be suppressed and the ions will begin to drift out of the center of the beam.  We model the process of increasing neutralization fraction during an electron bunch train using $f(t) = 1 - (1 - f_0)\exp(- t/ \tau_1)$, where $\tau_1$ is a time constant that defines the ion creation rate and $f_0$ is the initial neutralization fraction (which is not necessarily 0 in the steady state).  While the beam is off, the trapped ion density decays exponentially with a time constant $\tau_2$.  Figure \ref{figsawtooth} illustrates this creation and clearing process.

\begin{figure}[htb floatfix]
   \centering
   \includegraphics*[width=85mm]{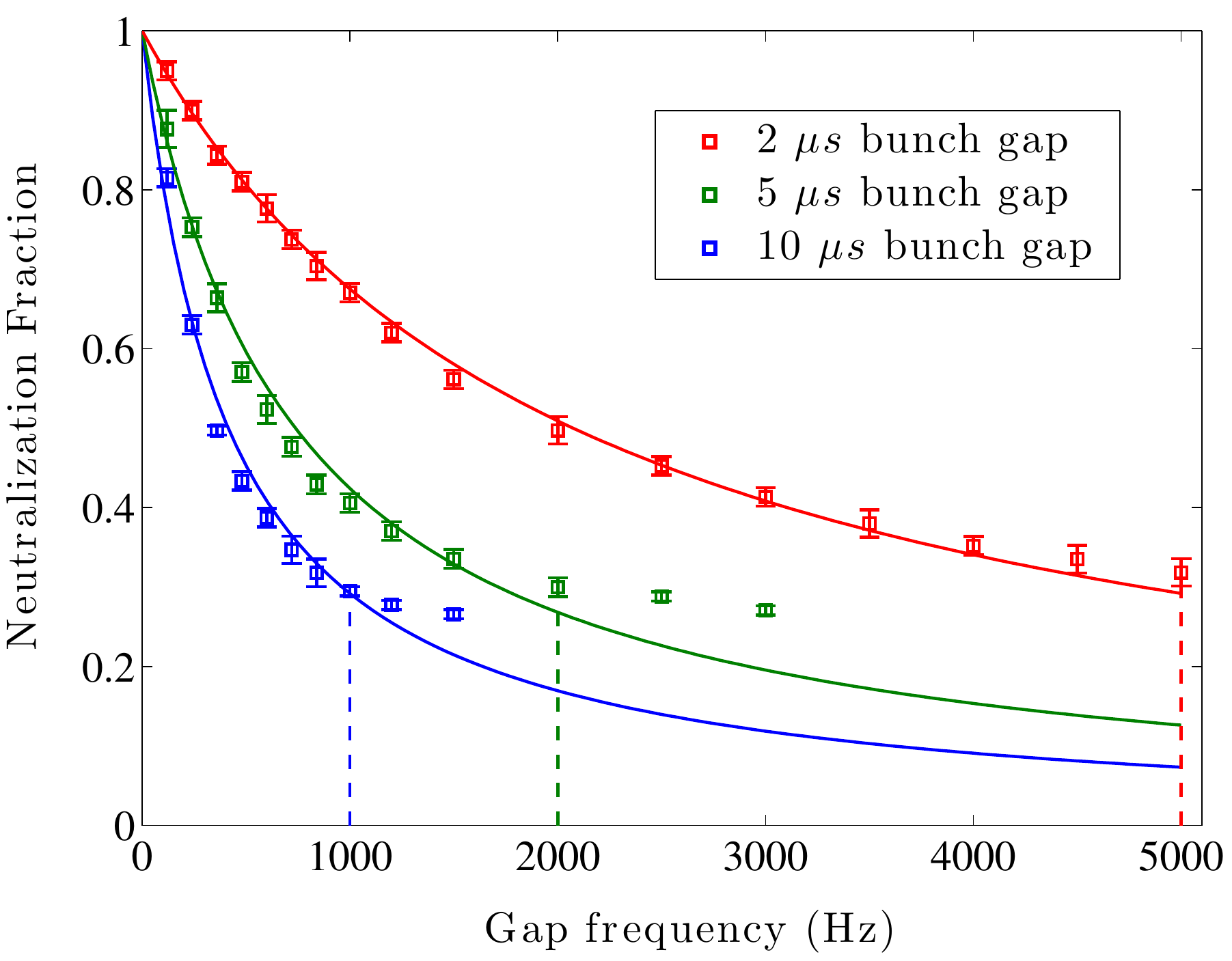}
   \caption{Increasing the frequency and duration of bunch gaps reduces the trapped ion density as shown by the residual ion current hitting a clearing electrode.  The curves are best fits obtained using equation (\ref{ridiculously_simplified_fraction}).  The vertical dotted lines mark the locations where a 1 percent reduction in beam current results in an approximately 70 percent reduction of trapped ion density.  }
   \label{figbunch_gaps2}
\end{figure}

The average value of the neutralization fraction (i.e. steady state solution) determines the amount of clearing we have observed experimentally.  From our model, the average ionization fraction is given by

\begin{equation}\label{ridiculously_simplified_fraction}
	f_{avg}(R_g) = \frac{1}{1 + \big( \frac{\tau_1}{\tau_2}\big)\big(\frac{R_g \Delta L_g}{1 - R_g \Delta L_g}\big)}
\end{equation}

\noindent
where $\tau_1$ is the characteristic creation time, $\tau_2$ is the characteristic clearing time, $R_g$ is the bunch gap frequency, and $\Delta L_g$  is the bunch gap duration.  Note that this is an approximate form, valid only for $(1/R_g - \Delta L_g)/\tau_1 \lesssim  0.5$ and $\Delta L_g/\tau_2 \lesssim  0.5$.  The full expression is derived in Appendix A.  The parameters $R_g$ and $\Delta L_g$ are well defined in the experiment, but the ratio of the two time constants $\tau_1$ and $\tau_2$ must be determined empirically from our data.  Our fit curves are compared with our data in Fig. \ref{figbunch_gaps2}, and the best fit ratio of $\tau_1/\tau_2$ for each curve is shown in Table \ref{table_ratio}.  

Assuming that $\tau_1$ is roughly the time it takes to achieve full beam neutralization (i.e. $\tau_1$ = $\tau_{create}$ = 5.2 ms), then the ratio $\tau_1/\tau_2$ can be used to predict a clearing time of 21 $\mu s$.  
This number is consistent with clearing rates estimated using the ion oscillation frequency (to be further explained in the next section).  For example, according to our data at 10 mA, a $\text{N}_\text{2}$ ion has an oscillation period of 17.5 $\mu s$, which is of the same order of magnitude as this estimate for $\tau_2$.   

Of particular note is that this simple model predicts the average amount of clearing depends only on the total time the beam is turned off ($R_g \Delta L_g$).  This observation agrees with our data.  For example, a 2 $\mu s$ gap at 5 kHz achieves the same amount of clearing as a 10 $\mu s$ gap at 1 kHz.  
In other words, the two data sets overlap when the horizontal axis is adjusted so that it becomes the total time the beam is turned off.
This suggests that the bunch gap method offers some flexibility, and may potentially allow the user to avoid RF beam loading problems by choosing the correct combination of gap duration and frequency.

Our data in Fig. \ref{figbunch_gaps2} shows that an approximately 70 percent reduction in ion density can be achieved while retaining nearly 99 percent of the maximum beam current.  To further reduce the number of trapped ions, one must increase the bunch gap frequency or duration, and introduce even more beam downtime.  According to our model, a 99 percent reduction in ion density would require over a 30 percent reduction in maximum beam current.  This is unacceptable for most ERL applications.  However, this is a large extrapolation of our model, and more data must be taken to determine the true limits of this clearing method.  The shortest possible gap that can still achieve clearing also has yet to be determined.  This is something we would like to examine in future experiments.  
\begin{table}[tb floatfix]
\caption{The ratio of creation to clearing times, found empirically from fits to our data.  }
\label{table_ratio}
\begin{tabular}{cc}
	\toprule
		\setlength\tabcolsep{12pt}
	\textbf{Gap duration ($\mu s$)                  } & \textbf{                        $\tau_1/\tau_2$}       \\
    	\hline
	2  &  2.4 $\times 10^{2}$ \\
	5  &  2.7 $\times 10^{2}$ \\
   	10 & 2.4 $\times 10^{2}$ \\
	\botrule	
\end{tabular}
\end{table}

%%%%%%%%%%%%%%%%%%%%%%%%%%%%%%%%%%%%%%%%%%%%%%%%%
%
%  Beam Shaking
%
%%%%%%%%%%%%%%%%%%%%%%%%%%%%%%%%%%%%%%%%%%%%%%%%%

\subsection{Beam Shaking}

In addition to their longitudinal drifting, the ions oscillate transversely in the beam's potential well.  One can imagine that the ion cloud and electron beam form a coupled oscillator.  By driving the beam at the trapped ions' oscillation frequency, a resonance is induced that kicks the ions out of the center of the beam.  This characteristic frequency should depend on the beam size and beam current.  

In order to determine the frequency of trapped ion oscillations inside an electron beam, we must first calculate the force acting on the ions.  The Coulomb force generated by an infinitely long, rotationally symmetric Gaussian beam can be derived using Gauss's  law, and is given by \cite{Keil}

\begin{equation}\label{beamforce}
	F(r) = \frac{\lambda e^2}{2 \pi \epsilon_0 r} \bigg{[} 1 - \exp\bigg{(}-\frac{r^2}{2 \sigma_b^2}\bigg{)} \bigg{]} 
\end{equation}

\noindent
where $r$ is the distance from the center of the beam, $\lambda$ is the number of electrons per unit length, and $\sigma_b$ is the rms width of the electron beam.  According to our simulations \cite{Colwyn}, the beam in the photoinjector is very nearly round for our experimental parameters, making this an appropriate approximation for our case. By linearizing this force, we are able to treat the ion's motion inside the beam as a simple harmonic oscillator.  The equation of motion in this case is then

\begin{equation}\label{SHO}
	\frac{d^2 r}{dt^2} + \omega_i^2 r = 0
\end{equation}

\noindent
where $\omega_i$ is the oscillation frequency of the ions.  Using the linearized form of (\ref{beamforce}), it follows that this oscillation frequency is given by \cite{chao}

\begin{equation}\label{ionfreq}
	\omega_i = \sqrt{\frac{2 r_p c}{e} \frac{I}{A \sigma_b^2}}
\end{equation}

\noindent
where $I$ is the beam current, $A$ is the atomic mass of the ion species,  and $r_p$ is the classical proton radius.  This formula can be used to estimate the frequency needed to clear out the ions.  Over the course of this experiment, we attempted to test the validity of this theory as well as the scaling laws it predicts.  

During this experiment the clearing electrode was used to shake the beam vertically.  It was placed approximately 1 m from the exit of the accelerating cavity.  A sinusoidally time varying voltage was applied to the electrode using a function generator and high voltage amplifier.  Oscillation frequencies were predicted to be in the 10-100 kHz range, so this is the primary range over which the experiment was performed.  

\begin{figure}[ht floatfix]
   \centering
   \includegraphics*[width=85mm]{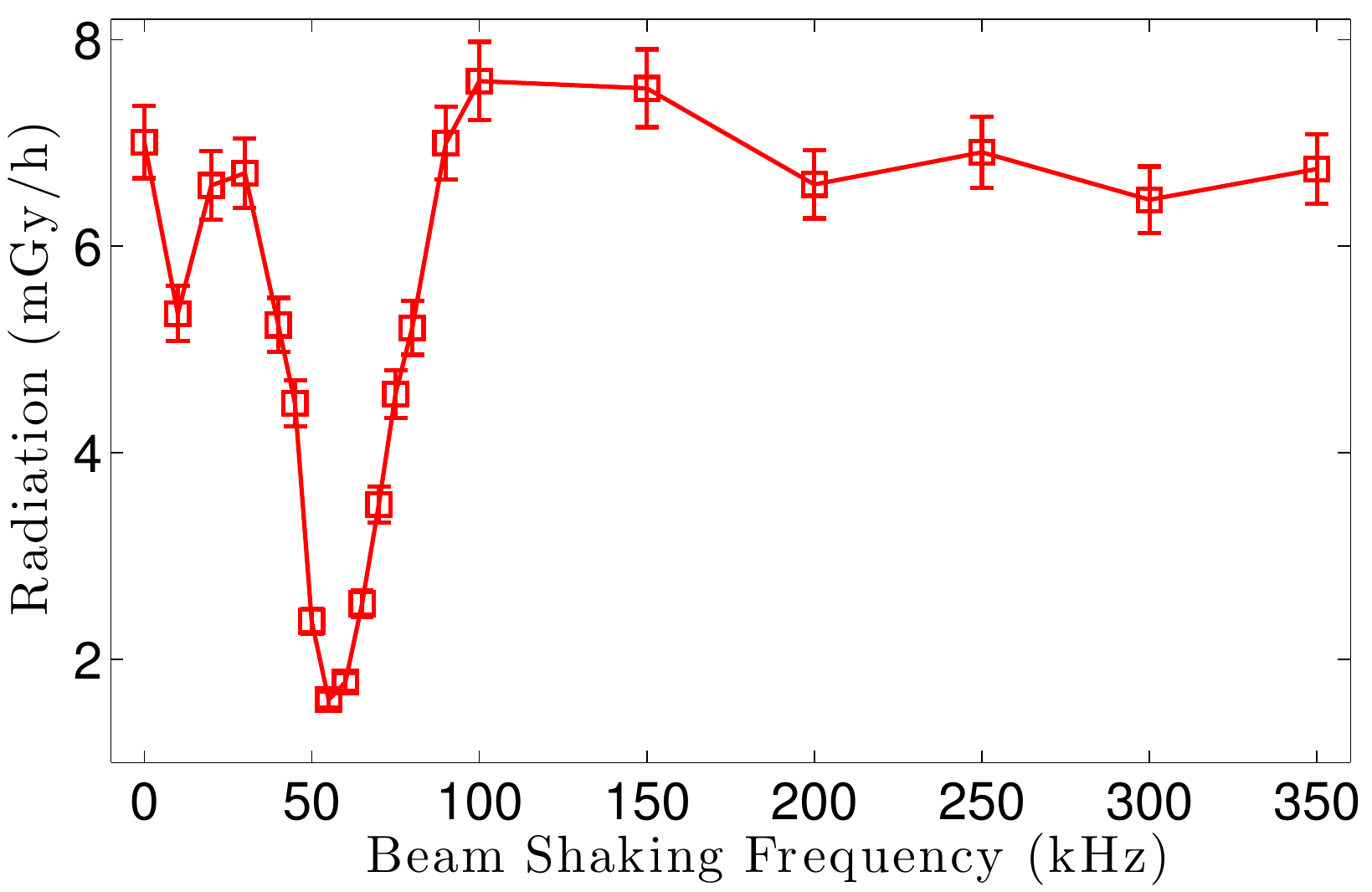}
   \caption{Shaking the beam at frequencies near the ion oscillation frequency eliminates the excess radiation caused by beam-ion interactions.}
   \label{measure1}
\end{figure}

Because our clearing electrode was being used to shake the beam, we could not measure the residual ion density using the picoammeter and electrode.  We were instead forced to rely solely on our indirect radiation measurements.  When the ions are cleared from the center of the beam pipe at resonance, the excess radiation caused by beam-ion collisions should vanish.  Thus, by measuring this radiation as a function of beam shaking frequency and noting the frequencies where the radiation vanishes, we are able to determine the oscillation frequencies of the ions.  The maximum voltage applied to the clearing electrode was adjusted as needed to completely clear the radiation at resonance, but the shaking amplitude never exceeded 0.5 mm for beam sizes of approximately 2--4 mm.  Generally results were visible for a shaking amplitude of roughly 0.1 mm.  An example of a typical measurement is shown in Fig. \ref{measure1}.  Measurements were taken for several gas species, including $\text{N}_\text{2}$, Ar and Kr, in order to confirm that the measured resonance frequency scaled correctly with ion mass.  

An attempt to shake the beam using broadband white noise was made, but this method did not result in any observable reduction in radiation.  At the present time, we do not understand why this method works in the MLS \cite{Feikes} and not the Cornell photoinjector.

\begin{figure}[ht floatfix]
   \centering
   \includegraphics*[width=85mm]{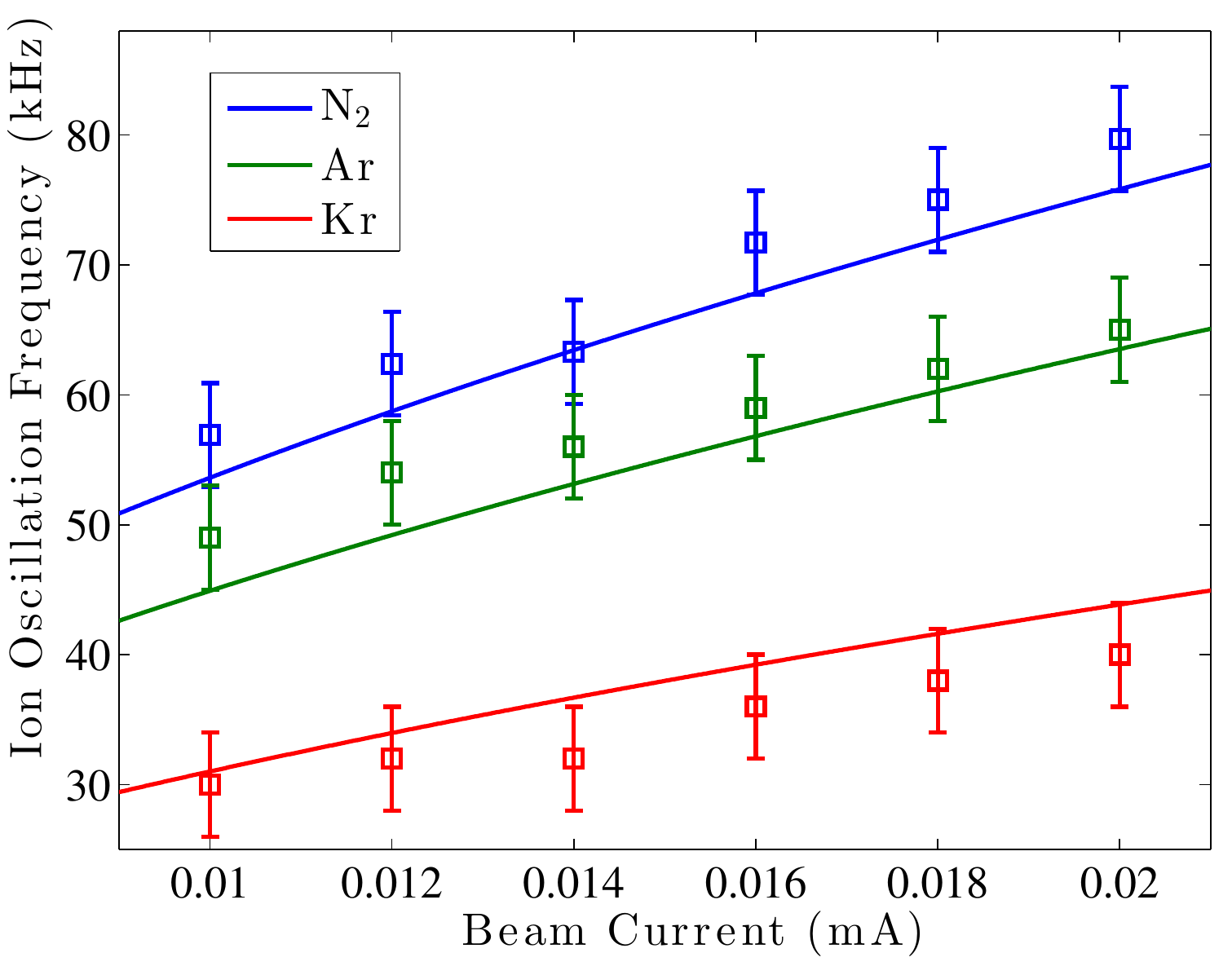}
   \caption{Resonance frequencies for various beam currents and ion species.  The circles represent data points, while the lines indicate best fits in the form of equation (\ref{ionfreq}), where the beam size is used as a fit parameter. }
   \label{figfreq}
\end{figure}

In the course of our experiments we attempted to confirm the three scaling laws predicted by equation (\ref{ionfreq}):  resonance frequency as a function of beam current, ion mass, and beam size.   Because the resonance peaks were quite broad, a fitting algorithm was used to fit the data, and the maximum value was taken as the resonance frequency.  
Figure \ref{figfreq} shows the measured resonance frequencies for beam currents over the range 10-20 mA, and for three different gas species.  Error bars for the data points are typically $\pm $ 3 kHz, and are due to systematic shifts in resonance frequency due to changing the electrode voltage, as well as statistical fluctuations. 

Here it is shown that the resonance frequency scales as predicted with beam current and ion mass.  This suggests that the resonance frequency required to clear the ions is indeed the ion oscillation frequency.    Given our lack of actual beam size measurements during this experiment, and the fact that GPT has not been experimentally verified in this beam parameter range, the beam size was treated as a fit parameter for our data.  A value of $\sigma_b$ = 4.2 mm was used to obtain the fit curves for the data in Fig. \ref{figfreq}.  GPT predicts a beam size of approximately 2 -- 3 mm between the clearing electrode and the gas leak, which is reasonably close to this fit value. 

\begin{figure}[ht floatfix]
   \centering
   \includegraphics*[width=85mm]{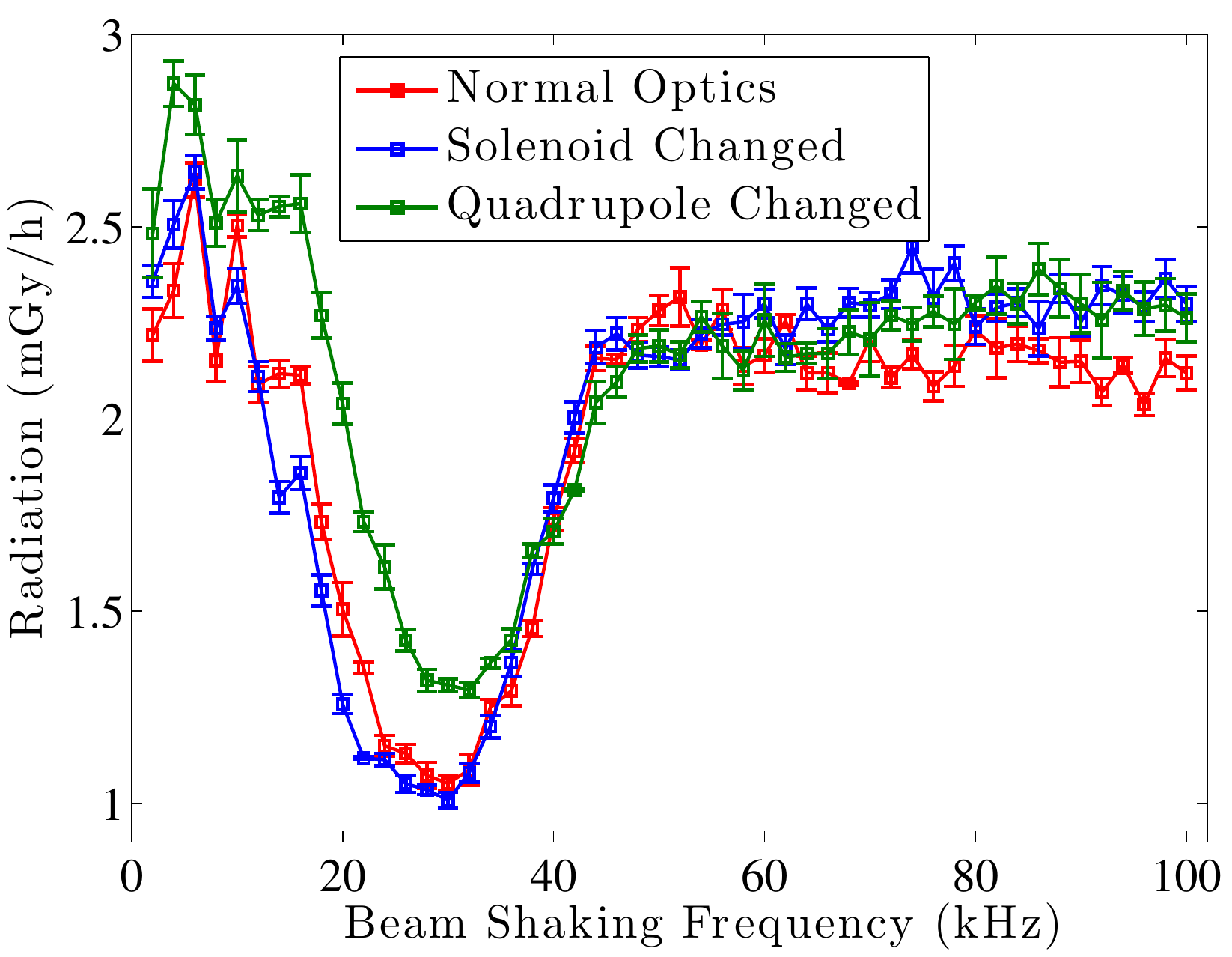}
   \caption{Radiation levels for various beam shaking frequencies.  Changing the beam size by over a factor of 2 does not result in a significant shift of the resonance frequency, in disagreement with theory.}
   \label{beamsize}
\end{figure}

However, the resonance frequency did not scale with beam size, as predicted by our theory.  Changing the beam size by almost a factor of 3 (using GPT simulations as a guide during operation) at the gas leak using a solenoid or a quadrupole magnet lead to a negligible change in resonance frequency, as shown in Fig. \ref{beamsize}.  

There are a few possible explanations.  This factor of 3 change in beam size was predicted using GPT, and it is possible that the beam size was not actually changing during the experiment.
Another possibility is that changing the optics settings only longitudinally moved the beam waist location while keeping the beam size constant at that waist.  Because the ions accumulate at beam size minima, the longitudinal location of the minima does not matter as much as the transverse beam size at that location.  However GPT does not predict this sort of behavior. 
In the future, repeating this experiment with beam profile diagnostics is necessary to determine why we were unable to observe a dependence on transverse beam size.

\section{Conclusions}

Experiments were performed to test the effectiveness of three different ion clearing techniques in the Cornell ERL photoinjector.  During the first experiment, we installed a DC clearing electrode and measured the amount of ions it removed as a function of clearing voltage.  It was found that the voltage necessary for maximum clearing can be predicted by considering the suppression of the transverse potential of a constant charge distribution beam.  The total measured ion current can also be used to estimate the effective longitudinal range of the clearing electrode.  In the second experiment, we introduced bunch gaps and determined the amount of clearing for different combinations of bunch gap duration and frequency.  The amount of clearing depended only on the total time the beam was turned off, and was independent of the bunch gap duration and frequency.  Finally, we used an electrode to shake the beam sinusoidally and resonantly clear out any trapped ions.  In this case, the shaking frequency needed to induce a resonance was the trapped ion oscillation frequency.  

From an ion mitigation standpoint, clearing electrodes appear to remain the most straightforward option.  A single electrode seems to clear most of the trapped ions in the photoinjector, especially because the region of interest is rather short (only about 5 m).  A larger accelerator would require the deployment of clearing electrodes near most beam size minima and other pockets of high ion concentration.  This may become difficult or expensive to implement in machines with low beta functions (on the order of m).  In this case, simulations must be done to determine the optimal placement of clearing electrodes \cite{Spethmann}.  The voltage required for full ion clearing can be predicted using a simple formula, and this can be used to design a proper clearing electrode.  For the photoinjector, the required voltage was rather small (28 V) compared to much higher energy accelerators which may require upwards of 1kV \cite{poncet1, Feikes}. This is due to a large difference in transverse beam size, which is typically mm in the photoinjector, as compared to beam sizes on the order of 10 $\mu m$ in storage rings or synchrotrons.  

In larger accelerators, beam shaking may be a more cost-effective option, because it only requires installing one or two electrodes to shake the beam.  The question remains whether or not the shaking amplitude is tolerable.  However, in practice, shaking appears to work for amplitudes that are much smaller than the transverse beam size.

The most promising results from these experiments are the bunch gap measurements.  Previously it was thought that this method was impossible in ERLs due to problems with RF beam loading \cite{Georg05}.  However, the size and frequency of the bunch gaps appears to be rather flexible, as the amount of clearing depends only on the total time the beam is turned off.  In addition, significant clearing can be achieved while retaining nearly 99 percent of the nominal beam current.  With this new information in mind, an analysis of beam loading and bunch gap mitigation in ERLs merits further study.  

In the future, we would like to continue these experiments with a new beam profile monitor capable of operating at high beam current \cite{wirescanner}.  This diagnostic is currently undergoing bench testing and should be available soon.  This will allow us to determine transverse beam sizes and supplement our results.  In addition, these measurements will be our first glimpse of any beam changes due to ions at high current in the photoinjector.  An attempt will also be made to take data for even higher beam currents above 20 mA.  

\section{acknowledgments}

The authors would like to thank Atoosa Meseck for helpful discussions concerning the planning of this experiment.  This work was supported by financial assistance from the U.S. Department of Energy (Grant No. DE-SC0012493) and the National Science Foundation (Award No. NSF-DMR 0807731).  

\appendix
\section{Average neutralization fraction}

\begin{figure}[htb floatfix]
   \centering
   \includegraphics*[width=85mm]{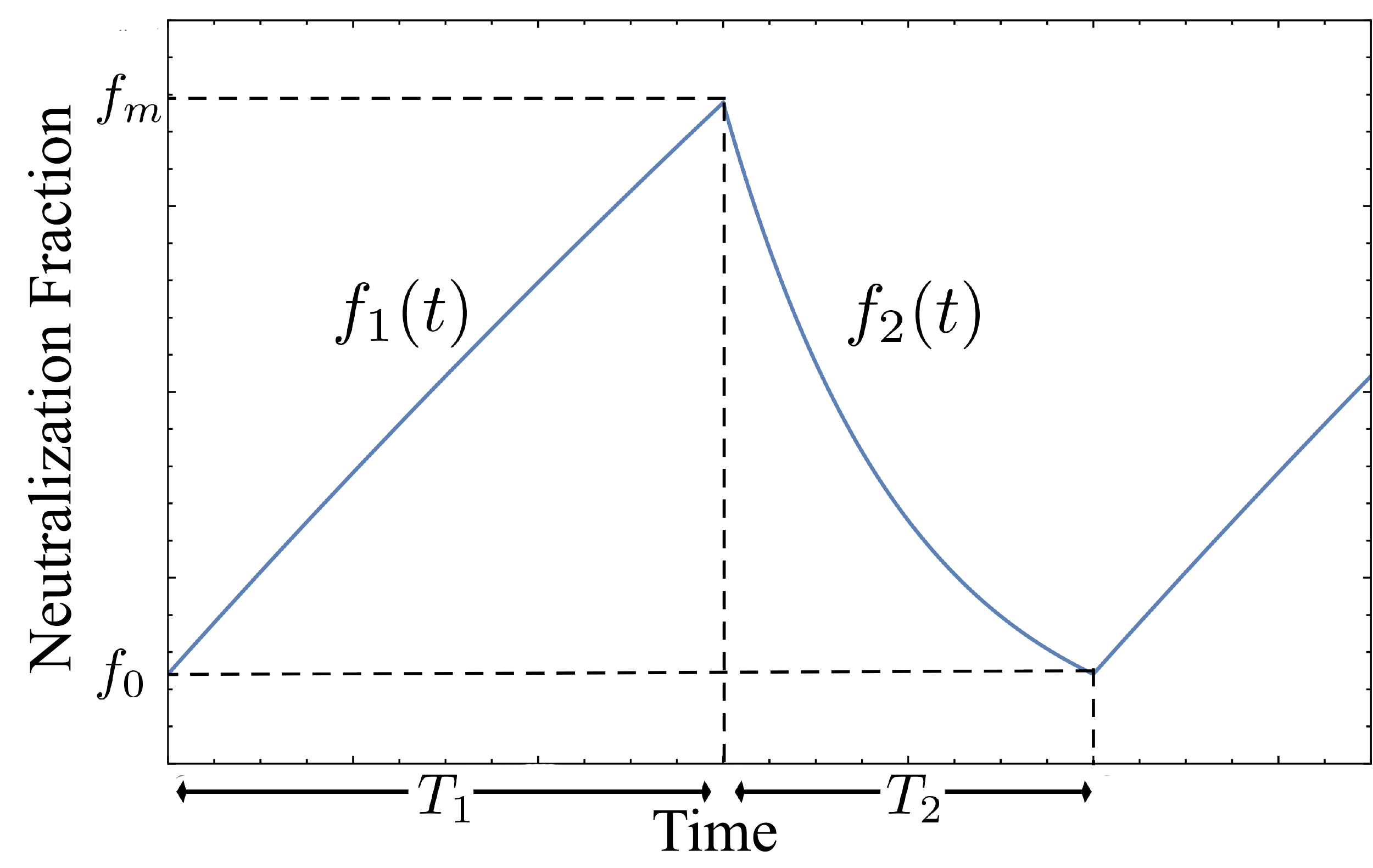}
   \caption{A sketch of the neutralization fraction while employing bunch gaps.  The neutralization fraction increases while the beam is on until some time $T_1$, then decays exponentially during the bunch gaps with a duration of $T_2$.  }
   \label{fig_derivation}
\end{figure}

To explain our bunch gap clearing data, we seek to obtain an expression for the average beam neutralization fraction as a function of bunch gap duration and frequency.  We begin by considering the creation rate of ions while the beam is turned on.  Ions are generated via collision ionization inside of the beam at a constant rate, therefore the ion density increases linearly with time.  However, the neutralization fraction can never exceed 1, because the positive ions will eventually screen the electron beam's negative potential and the ions will begin to drift out of the center of the beam pipe.   We model this behavior by assuming the neutralization fraction has a functional form of 

\begin{equation}\label{ion_creation_rate}
	f_{1}(t) = 1 - (1 - f_0) e^{-t/\tau_1}
\end{equation}

\noindent
where $\tau_1$ is a time constant that defines the creation rate and $f_0$ is the initial neutralization fraction (which is not necessarily 0 in the steady state).  This expression applies only while the beam is on, up until some time $T_1$,  as shown in Fig. \ref{fig_derivation}.  

During the bunch gaps, the neutralization fraction decays exponentially  as given by

\begin{equation}\label{ion_clearing_rate}
	f_{2}(t) = f_m e^{-(t - T_1)/\tau_2}
\end{equation}

\noindent
where $\tau_2$ is the characteristic clearing time, and the constant $f_m$ is the maximum neutralization fraction reached in the steady state (because $f_2(T_1) = f_m$ in Fig. \ref{fig_derivation}).  Using this, along with the fact that the total function must be continuous at both $f_1(T_1 + T_2) = f_2(T_1 + T_2) = f_0$ and $f_1(T_1) = f_2(T_1) = f_m$, one can obtain expressions for both $f_0$ and $f_m$, 

\begin{equation}\label{f0}
	f_0 = \frac{1 - e^{\frac{T_1}{\tau_1}}}{1 - e^{\frac{T_1}{\tau_1} + \frac{T_2}{\tau_2}}},
\end{equation}

\begin{equation}\label{fm}
	f_m = 1 - (1 - f_0) e^{-T_1/\tau_1}.
\end{equation}

\noindent
Now that $f_1(t)$ and $f_2(t)$ are well defined, the average ionization fraction can be found via integration over the full time interval $T_1 + T_2$.  This is given by

\begin{equation}\label{integration}
f_{avg} = \frac{1}{T_1}\int_0^{T_1} f_1(t)\: \:+ \: \: \frac{1}{T_2}\int_{T_1}^{T_1 + T_2}f_2(t).
\end{equation}

\noindent
Performing this integral yields the formula for the average neutralization fraction

\begin{equation}\label{avg_ion_fraction}
f_{avg}  = \frac{T_1}{T_1 + T_2}-2 \frac{\tau_1 - \tau_2}{T_1 + T_2}\frac{\text{sinh}\bigg(\frac{T_1}{2 \tau_1}\bigg) \text{sinh}\bigg(\frac{T_2}{2 \tau_2}\bigg)}{ \text{sinh}\bigg(\frac{T_1}{2\tau_1}+\frac{T_2}{2\tau_2}\bigg) }.
\end{equation}

\noindent
This expression can be simplified for small $T_1/\tau_1$ and $T_2/\tau_2$ by using the approximation $\text{sinh}(x) \approx x$.  This ultimately yields an approximate form of the average neutralization fraction given by

\begin{equation}\label{ridiculously_simplified_fraction2}
	f_{avg} = \frac{1}{1 + \big( \frac{\tau_1}{\tau_2}\big)\big(\frac{T_2}{T_1}\big)}.
\end{equation}

\noindent
This approximation works well for $T_1/\tau_1 \lesssim  0.5$ and $T_2/\tau_2 \lesssim  0.5$.  By substituting $T_2 = \Delta L_g$ for the bunch gap duration and $T_1 = 1/R_g - T_2$ for the time the beam is on, one can obtain the expression found in equation (\ref{ridiculously_simplified_fraction}).

\bibliography{Detection_and_clearing_of_trapped_ions}
\bibliographystyle{h-physrev}

\end{document}